\documentclass{openjournal}

\usepackage{lipsum}
\usepackage{multirow}
\usepackage{amsmath}

\usepackage{xcolor}
\usepackage{textgreek}
\usepackage[utf8]{inputenc}
\usepackage[english]{babel}

\usepackage{hyperref}
\hypersetup{
    unicode, 
    colorlinks=true,
    linkcolor=linkcolor,
    citecolor=linkcolor,
    filecolor=linkcolor,
    urlcolor=linkcolor,
}
\usepackage{color,colortbl}
\definecolor{linkcolor}{rgb}{0.0,0.3,0.5}
\usepackage{tensind}
\tensordelimiter{?}
\DeclareGraphicsExtensions{.bmp,.png,.jpg,.pdf}
\usepackage{verbatim}
\usepackage[normalem]{ulem}
\usepackage{orcidlink}
\usepackage{soul}

\usepackage{xspace}

\newcommand{\LCDM}{$\Lambda$CDM\xspace}

\graphicspath{ {./figures/} }

\begin{document}
\title{Explaining Neural Networks on the Sky: \\
Machine Learning Interpretability for CMB Maps}

\author{Indira Ocampo\orcidlink{0000-0001-7709-1930}}
\email{indira.ocampo@csic.es}
\affiliation{Instituto de F\'isica Te\'orica UAM-CSIC}
\affiliation{Universidad Aut\'onoma de Madrid, Cantoblanco, 28049 Madrid, Spain}

\author{Guadalupe Cañas-Herrera\orcidlink{0000-0003-2796-2149}}
\email{canasherrera@strw.leidenuniv.nl}
\affiliation{Leiden Observatory, Leiden University, PO Box 9506, Leiden 2300 RA, The Netherlands}

\begin{abstract}
We present a framework for cosmological model selection using Neural Networks (NNs) trained directly on simulated Cosmic Microwave Background (CMB) temperature and polarisation maps. We also apply SHAP (Shapley Additive exPlanations) as an interpretability approach to map the network's decision-making process. By introducing a galactic mask as a diagnostic control, we obtain attribution maps that correctly exclude this region and confirm that the classification is driven exclusively by the unmasked sky. Ultimately, these maps verify that the architecture distinguishes between \LCDM and the non-standard feature model by evaluating the global statistical variance distributed across the valid cosmological regions.
We describe the generation of Planck-like CMB maps, and the implemented hybrid architecture that combines principal component analysis and NNs, optimised for classification tasks.
This work serves as a follow-up to previous analyses at the level of summary statistics and as a proof-of-concept for using machine learning and interpretability to evaluate the global statistical properties of masked CMB data. This diagnostic Machine Learning framework has the potential to enhance future detection of nontrivial inflationary signals and improve cosmological model discrimination, while also
introduces the Open Science project \texttt{SkyExplain}\footnote{\url{https://github.com/skyexplain}}, providing public access to the full pipeline for simulation, training, and interpretability of CMB map-based neural networks.
\end{abstract}


\maketitle

\section{Introduction}
\label{sec:intro}
The Cosmic Microwave Background (CMB) remains our most definitive window into the early Universe, providing a snapshot of the primordial fluctuations present at the epoch of recombination ($z \approx 1100$) \citep{chang2022snowmass2021, baumann2009tasi}. While the standard \LCDM model offers a robust description of these temperature and polarisation anisotropies, persistent tensions with late-time probes suggest that our understanding of the fundamental physical processes may be incomplete \citep{gu2025dynamical}. This has motivated the interest in beyond-\LCDM scenarios, particularly those involving localised or oscillatory features in the primordial power spectrum, which can arise from complex inflationary dynamics such as sharp turns in field space or multi-field interactions \citep{benetti2016bayesian}.

Traditionally, CMB analysis relies on the angular power spectra ($C_{\ell}^{TT}, C_{\ell}^{TE}, C_{\ell}^{EE}$). In our previous work \citep{ocampo2025neural}, we demonstrated that machine learning (ML) can effectively classify cosmological models using these 1D harmonic-space representations. We also introduced one of the first studies of interpretability in ML, which allowed us to look beyond the classification performance and trace how specific features in the power spectra drive the model's decisions. This work aims to complement our previous classification framework to operate directly on simulated Planck-like maps. 
At the level of the maps, inference is often applied to detect the non-Gaussian or spatially localised imprints predicted by theoretical extensions to \LCDM \citep{chandra2022investigating}, while also providing methodological advantages for handling observational constraints.

This transition to the direct analysis of CMB maps introduces specific observational and computational challenges, primarily the need to process a large volume of pixels while accounting for instrumental noise and galactic contamination. As a result of the increase in requirements of computational tools to handle these data, ML has gained significant attention. For instance, deep learning pipelines were applied to perform scalable foreground cleaning tasks \cite{amato2025cmb}, and to isolate complex, non-Gaussian features at the pixel level, such as detecting the signatures of cosmic strings \cite{ciuca2019convolutional}. Moreover, the potential of map-level ML for detecting departures from statistical isotropy has been recently demonstrated in works such as \cite{noltmann2026cosmic, tamosiunas2024cosmic}.

From an observational point of view and in standard harmonic-space analysis, applying a galactic mask leads to the mixing of multipole modes. This requires specific mathematical corrections to estimate the corresponding power spectrum. The direct analysis at the map level avoids this reconstruction step, allowing the model to evaluate the available sky exactly as it is observed \cite{costanza2024enhancing}.

A significant challenge in applying ML to cosmology is the difficulty in mapping the model's internal representations back to the physical mechanisms, i.e. the lack of interpretability. To make sure a model is actually finding physical signatures, we need to be able to understand how it makes its decisions. Recent advancements in applying ML interpretability techniques to cosmological data include studies by \cite{piras2025lambda, ntampaka2022importance, ocampo2025enhancing}. The goal of this study is to train a pipeline based on NNs to discriminate CMB temperature and polarisation maps simulated using two different models: \LCDM and an alternative non-standard primordial feature model. We include a galactic mask in the simulated data to replicate realistic sky-cut conditions, and to evaluate that the neural network is correctly focusing on the unmasked cosmological regions. For this, We implement the Shapley Additive Explanations (SHAP) package \cite{Lundberg2017SHAP}, a post-hoc interpretability method, i.e. a technique applied after training to see the influence of the input data on the architecture’s predictions. By applying SHAP to this context, we generate attribution maps that allow us to visualise which spatial regions contribute most to the classification. We deliberately include the galactic mask to evaluate the model's reliability. Because this region contains no physical data, it allows us to unambiguously verify that the model ignores the masked area and focuses exclusively on the unmasked sky. This represents a novel application of interpretability methods at the CMB map level, offering a diagnostic tool to ensure that high classification performance reflects actual cosmological information.

Although the simulated maps generated for this study are Gaussian random fields, applying a map-level analysis here serves a specific, novel diagnostic purpose. The main contribution of our methodology is the integration of a spatial interpretability framework alongside the galactic mask. By training the NN on incomplete sky maps, we can apply interpretability tools as a diagnose of its decision-making process. This allows us to explicitly verify that the network successfully identifies the global statistical variance within the valid cosmological regions. Confirming this reliability assessment on well-understood Gaussian fields provides a tested pipeline, this will allow us to validate the network's diagnostic transparency, and set a framework for future studies of higher-order imprints.

This paper is organised as follows: \autoref{sec:featuremodel} introduces the primordial feature model under consideration. \autoref{sec:methodology} details the simulation of Planck-like maps (including masking), the preprocessing pipeline, and the NN architecture. Our results, including a comparative performance analysis and an exploration of model interpretability, are presented in \autoref{sec:results}, where we also show attribution maps of the most informative regions in pixel space that correspond to those of the simulated CMB ones. Finally, we discuss the broader implications of map-level inference, conclusions and future directions in \autoref{sec:conclusions}.\\

\section{Theoretical predictions for primordial features}\label{sec:featuremodel}

The observed homogeneity and isotropy of the Universe suggest that primordial density perturbations were nearly Gaussian and scale-invariant, a result naturally explained by the inflationary paradigm. In the standard single-field slow-roll scenario, the primordial power spectrum of comoving curvature perturbations is typically parametrised by a simple power law:
\begin{equation}
P_{\mathcal{R}, 0}(k)=\frac{2 \pi^2}{k^3} \mathcal{P}_{\mathcal{R}, 0}(k)= A_{\mathrm{s}}\left(\frac{k}{k_{\star}}\right)^{n_{\mathrm{s}}-1},
\label{eq:PPs}
\end{equation}
where $A_{\mathrm{s}}$ is the amplitude, $n_{\mathrm{s}}$ is the spectral index, and $k_{\star} = 0.05\, \mathrm{Mpc}^{-1}$ is the chosen pivot scale. 

Although this vanilla \LCDM description is remarkably consistent with current data, it remains a phenomenological fit that may simplify the complex physical processes of the inflationary era \citep{lodha2023searching, kamerkar2023machine}. In this sense, \autoref{eq:PPs} is compatible with the assumptions of the \LCDM, setting a phenomenological parametrisation of an \textit{almost}-scale invariant power spectrum. However, it lacks an intrinsic mechanism to explore further inflationary scenarios or other alternatives, making the study of deviations from this power-law a trending topic in cosmology \citep{features_review}. Deviations from this smooth power law, collectively known as primordial features, offer an interesting window into the fundamental dynamics of the early Universe. These features, which can manifest as localised sharp steps or global oscillations, would provide a signature of physics beyond the simplest slow-roll models. These deviations are modeled as perturbations overlaid upon the standard primordial power spectrum,
\begin{equation}
P_{\mathcal{R}}(k)=P_{\mathcal{R}, 0}(k)\left[1+ \frac{\Delta P_{\mathcal{R}}}{P_{\mathcal{R}, 0}}\left(k\right)\right].
\label{eq:PPs_feature}
\end{equation}
In this work, we employ a template where these features appear as periodic fluctuations in Fourier space, maintaining a constant amplitude relative to the standard slow-roll prediction. The specific template is given by,
\begin{equation}
\frac{\Delta P_{\mathcal{R}}}{P_{\mathcal{R}, 0}}=A_{\mathrm{lin}} \sin \left(\omega_{\mathrm{lin}} \frac{k}{k_{\star}} +\phi\right),
\label{eq:Feature}
\end{equation}
where $A_{\mathrm{lin}}$ is the oscillation amplitude, $\omega_{\mathrm{lin}}$ is the frequency, and $\phi$ is an arbitrary phase. Following the fiducial values explored in the context of future surveys like Euclid \citep{ballardini2024euclid}, we adopt:
\begin{equation}
\Theta_{\text {lin }}=\left\{{A}_{\text {lin }}=0.01, \omega_{\text {lin }}=10, \phi_{\mathrm{lin}}=0\right\} .
\label{eq:lin_oscillations}
\end{equation}
To test the robustness of our map-based NNs, we vary $A_{\mathrm{lin}}$ and $\omega_{\mathrm{lin}}$ around these fiducial values, ensuring our simulations remain consistent with current Planck 2018 constraints \citep{Planck:2018jri}. Notably, the \LCDM cosmology is nested within this framework as the limit where $A_{\mathrm{lin}} \to 0$ and $\omega_{\mathrm{lin}} \to 0$. Rather than explicitly reconstructing the angular power spectrum, we aim to train our model to evaluate the global statistical shifts introduced by these subtle oscillatory features directly from the CMB maps. Building upon these classification efforts, our complementary work introduces a diagnostic framework. This aims to deliver importance attribution maps that verify the network is correctly evaluating the global statistical properties of the unmasked CMB sky to detect non-standard linear features.

Detecting such oscillatory signatures would fundamentally refine our constraints on the early Universe and the mechanism driving inflation. While, several feature templates have been tested against Planck 18 data \citep{Planck:2018jri}, an independent pipeline is needed to probe the large-scale structure (LSS) signals present in upcoming photometric and spectroscopic surveys. This enhances the sensitivity to high-frequency signals and complements CMB measurements.

\subsection{CMB polarisation and Stokes parameters}
\label{subsec:stokes}
In addition to temperature anisotropies, the CMB exhibits a small but measurable degree of linear polarisation. This polarisation is generated through Thomson scattering of radiation off free electrons in the presence of a local quadrupole anisotropy in the photon distribution function \cite{durrer2020cosmic}. The radiation field is described in terms of the Stokes parameters $T, Q$, and $U$, defined as functions on the sphere. The temperature anisotropy $T(\hat{n})$ is a scalar field, whereas $Q(\hat{n})$ and $U(\hat{n})$ characterise linear polarisation and depend on the choice of local orthonormal basis perpendicular to the line of sight $\hat{n}$ .

In a Cartesian basis, orthogonal to the line of sight, and for a monochromatic plane wave propagating along $\hat{z}$, we define the time-averaged electric field components $E_x(t)$ and $E_y(t)$ and write the Stokes parameters as
\begin{equation}
\begin{aligned}
I &= \big\langle |E_x|^2 \big\rangle + \big\langle |E_y|^2 \big\rangle, \\
Q &= \big\langle |E_x|^2 \big\rangle - \big\langle |E_y|^2 \big\rangle, \\
U &= 2\,\mathrm{Re}\big\langle E_x E_y^{\ast} \big\rangle, \\
V &= 2\,\mathrm{Im}\big\langle E_x E_y^{\ast} \big\rangle.
\end{aligned}
\label{eq:stokes-def}
\end{equation}
Here $I$ denotes the total intensity, $Q$ and $U$ describe linear polarisation, and $V$ describes circular polarisation \citep{HuWhite1997}. The temperature map $T(\hat{n})$ corresponds to intensity fluctuations, $T \propto I - \bar{I}$ \citep{Gorski2005, aghanim2020planck}.
Thomson scattering does not produce circular polarisation in the standard scenario, so $V$ is expected to vanish to an excellent approximation \citep{Kosowsky1996}.
Throughout this work we therefore focus on the triplet $(T,\;Q,\;U)$.

On the sphere, $Q(\hat{n})$ and $U(\hat{n})$ are not scalar quantities: under a rotation of the local basis by an angle $\varphi$, they transform as a spin-2 field \citep{Zaldarriaga1997,Kamionkowski1997}.
This property is captured by the combinations
\begin{equation}
(Q \pm i U)(\hat{n})=\sum_{\ell=2}^{\infty} \sum_{m=-\ell}^{\ell} a_{\ell m}^{\pm 2} \, {}_{\pm 2}Y_{\ell m}(\hat{n}).
\end{equation}
where ${}_{\pm 2}Y_{\ell m}$ are spin-weighted spherical harmonics and $a_{\ell m}^{\pm 2}$ are the corresponding harmonic coefficients \citep{Kamionkowski1997}.
One then defines scalar $E-$ and pseudo-scalar $B-$mode coefficients by
\begin{equation}
a_{\ell m}^{E} = -\frac{1}{2} \left( a^{(2)}_{\ell m} + a^{(-2)}_{\ell m} \right), 
\qquad
a_{\ell m}^{B} = -\frac{i}{2} \left( a^{(2)}_{\ell m} - a^{(-2)}_{\ell m} \right),
\label{eq:EB-def}
\end{equation}
which are invariant under rotations of the local basis \citep{Zaldarriaga1997}. Scalar (density) perturbations generate the observed temperature variations $T$ and $E$ modes but no $B$ modes at linear order, whereas tensor perturbations and weak gravitational lensing produce both $E$ and $B$ structures \citep{keating2006polarization}. This distinction makes polarisation a sensitive probe of the physics of the early Universe \citep{quiet2012second}.

Assuming statistical isotropy and Gaussianity, the statistical properties of the CMB are fully characterised by the angular power spectra
\begin{equation}
C_\ell^{XY} = \langle a_{\ell m}^X a_{\ell m}^{Y*} \rangle,
\end{equation}
where $X, Y \in \{T, E, B\}$, and the expectation value is taken over an ensemble of realisations. For a more detailed review, we refer the reader to \cite{durrer2020cosmic}.

In practice, observations provide maps of the Stokes parameters $T$, $Q$, and $U$, from which the harmonic coefficients $a_{\ell m}^T$, $a_{\ell m}^E$, and $a_{\ell m}^B$ are reconstructed. The angular power spectra $C_\ell^{XY}$ are then estimated from these coefficients, forming the basis for cosmological parameter inference. This establishes the direct link between the observed sky maps and the statistical quantities used to constrain cosmological models \cite{sullivan2025methods}.

However, calculating isotropic $C_\ell$ spectra averages out localised spatial information. In our simulations, we avoid this harmonic-space compression and work directly with maps of the Stokes parameters $(T,Q,U)$ in the HEALPix representation \citep{Gorski2005}, generated from the corresponding theoretical power spectra. To mimic realistic observational conditions and provide a benchmark for our interpretability framework, we apply a standard galactic mask to these simulated maps. While traditionally, masking is implemented to remove contaminated pixels from the $C_\ell$ estimation, retaining the masked map structure in pixel space allows us to physically validate our ML model. As discussed in the previous section, by analysing the maps directly, we can verify whether the our architecture correctly ignores the unphysical, masked regions and instead extracts its cosmological information from the unmasked sky.\\

\section{Methodology}\label{sec:methodology}
Our analysis pipeline follows a sequence of phases: preparation of spherical temperature and polarisation CMB maps, dataset generation, data pre-processing, ML architecture design and training, performance assessment, and post-hoc ML interpretability. Each step is described in detail in the sections that follow. The full implementation is publicly available via the \texttt{SkyExplain} GitHub organisation\footnote{\url{https://github.com/SkyExplain}}, with dedicated \texttt{python} packages for \texttt{SkySimulation}, \texttt{SkyNeuralNets}, and \texttt{SkyInterpret}.

\subsection{Simulation of Planck-like Maps} \label{subsec:SimulatedData}
We generated full-sky CMB temperature and polarisation maps with Planck-like instrumental characteristics, adopting angular resolution, beam smoothing, and noise levels consistent with the performance of the \textit{Planck} satellite mission \cite{planck2018_overview}.
First, the theoretical input angular power spectra $C_{\ell}^{TT}$, $C_{\ell}^{TE}$ and $C_{\ell}^{EE}$ were computed with the Boltzmann Solver \texttt{CAMB} (version 1.5.9) \citep{lewis2011camb, lewis2000efficient} for both the baseline \LCDM model and the representative feature model. In our simulations we set $C_{\ell}^{BB}=0$, reflecting the fact that Planck has not detected a primordial $B-$mode signal and that the lensing contribution is small compared with the temperature and $E-$mode spectra \citep{Zaldarriaga1997, HuWhite1997}. The oscillatory features in the primordial power spectrum discussed in~\autoref{sec:featuremodel} induce correlated oscillations in these spectra, which modify the joint statistical properties of the $(T,Q,U)$ fields on the sphere \citep{features_review,Hu2004}.

Planck uncertainties were then incorporated into these spectra\footnote{\label{fn:esa_archive}\url{https://pla.esac.esa.int/home}}, which were subsequently used as input to the \texttt{healpy} routine (version 1.18.1) to generate Gaussian random $(T,Q,U)$ maps on the sphere, by drawing spherical harmonic coefficients $a_{\ell m}$ consistent with the input $C_{\ell}$. This procedure naturally incorporates cosmic variance, i.e. the statistical scatter expected from different realisations of the same cosmological model. The Planck parameters held fixed in the simulations are listed in \autoref{tab:fixed}, while the ranges for the parameters that were varied are given in \autoref{tab:prior}.

\begin{table}[h!]
\centering
\begin{tabular}{|c|c|c|}
\hline
\textbf{Parameter} & \textbf{Description} & \textbf{Planck alone value} \\
\hline
\hline
$H_0$ [km s$^{-1}$ Mpc$^{-1}$] & Hubble constant & 67.32 \\
$\tau$ & Optical depth to reionisation & 0.0543 \\
$\Omega_k$ & Curvature density & 0 \\
$\sum m_\nu$ [eV] & Neutrino masses & 0.06 \\
\hline
\end{tabular}
\caption{Fixed Planck cosmological parameters used in the simulations of CMB maps.}
\label{tab:fixed}
\end{table}

We additionally applied a galactic mask (also available at the ESA archival data\textsuperscript{\ref{fn:esa_archive}}) to the simulated maps in order to mimic the incomplete sky coverage of real CMB observations. In practice, foreground emission from the Milky Way, primarily synchrotron, free–free, and thermal dust radiation, dominates over the cosmological signal near the Galactic plane, making those regions difficult for precision cosmological analyses \citep{planck2018_foregrounds}. Masking these contaminated areas produces maps with anisotropic sky coverage and mode coupling in harmonic space, thereby altering the statistical properties of the field and preventing direct recovery of full-sky summary statistics such as the angular power spectrum without dedicated estimators \citep{hivon2002master}. Moreover, the mask was also useful to verify whether the our architecture correctly learns the informative regions from the unmasked sky, this analysis will be discussed in the interpretability part.

\begin{table}[h!]
\centering
    \begin{tabular}{|c|c|c|c|c|c|c|}
    \hline
        \multirow{2}{*}{\textbf{Model}} & \multicolumn{6}{c|}{\textbf{Simulation (prior) range parameters}}\\
        \cline{2-7}
        & \textbf{$\omega_{\mathrm{cdm}}$} & \textbf{$\omega_{\mathrm{b}}$} & \textbf{$A_{\mathrm{s}}$} & \textbf{$n_{\mathrm{s}}$} & \textbf{$A_{\mathrm{lin}}$} & \textbf{$\omega_{\mathrm{lin}}$}\\
        \hline
        \hline
        \textbf{$\Lambda$CDM} & \multirow{2}{*}{[0.1, 0.15]} & \multirow{2}{*}{[0.021, 0.023]} & \multirow{2}{*}{[1.8, 2.4]$\times 10^{-9}$} & \multirow{2}{*}{[0.94, 0.99]} & - & - \\
        \cline{1-1}
        \cline{6-7}
        \textbf{Feature} & & & & & [0.01, 0.06] & [5, 25] \\
        \hline
    \end{tabular}
    \caption{Prior range of the varied cosmological parameters for the simulated CMB maps.}
    \label{tab:prior}
\end{table}

CMB temperature maps are naturally defined on the sphere and stored in the HEALPix format \citep{gorski2005healpix}, which represents the data as a one-dimensional array of pixels corresponding to equal-area regions on the sphere.

To mimic observational effects, the maps were convolved with a Gaussian beam corresponding to Planck’s effective angular resolution (FWHM $\sim$5 arcmin for high-frequency channels \citep{aghanim2020planck}) and noise realisations were added using a simulated Planck covariance matrix drawn from the Planck uncertainties \citep{aghanim2020planck}, thereby including anisotropic noise.

\begin{figure}[h!]
    \centering
    \includegraphics[width=\textwidth]{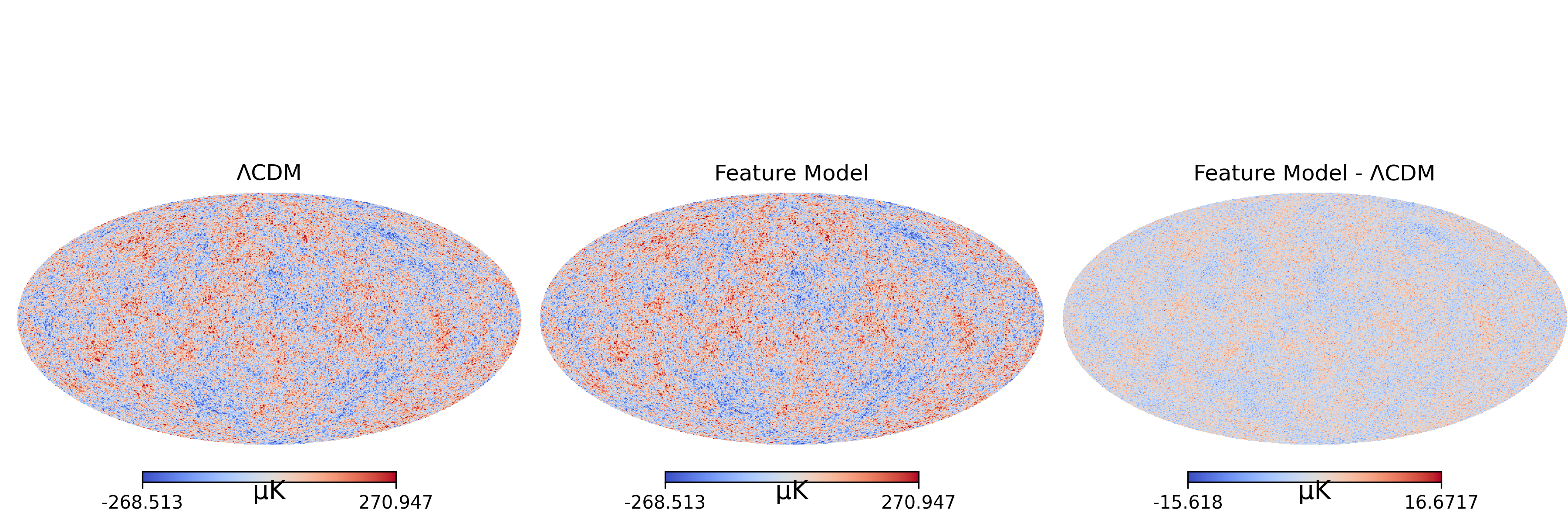}
    \caption{Simulated CMB temperature maps for the \LCDM model (left) and the primordial feature model (middle), along with their difference (right), for an amplitude of $A_{\mathrm{lin}} = 0.1$. The similarity between the individual maps reflects the dominance of primary CMB fluctuations, whereas the residual map reveals a coherent, spatially correlated signal associated with the primordial feature.} 
    \label{fig:diffmaps}
\end{figure}

\begin{figure}[!htbp]
  \centering
  \def\panelheight{0.19\textheight}
  \begin{minipage}[t]{0.45\textwidth}
    \centering
    \includegraphics[height=\panelheight]{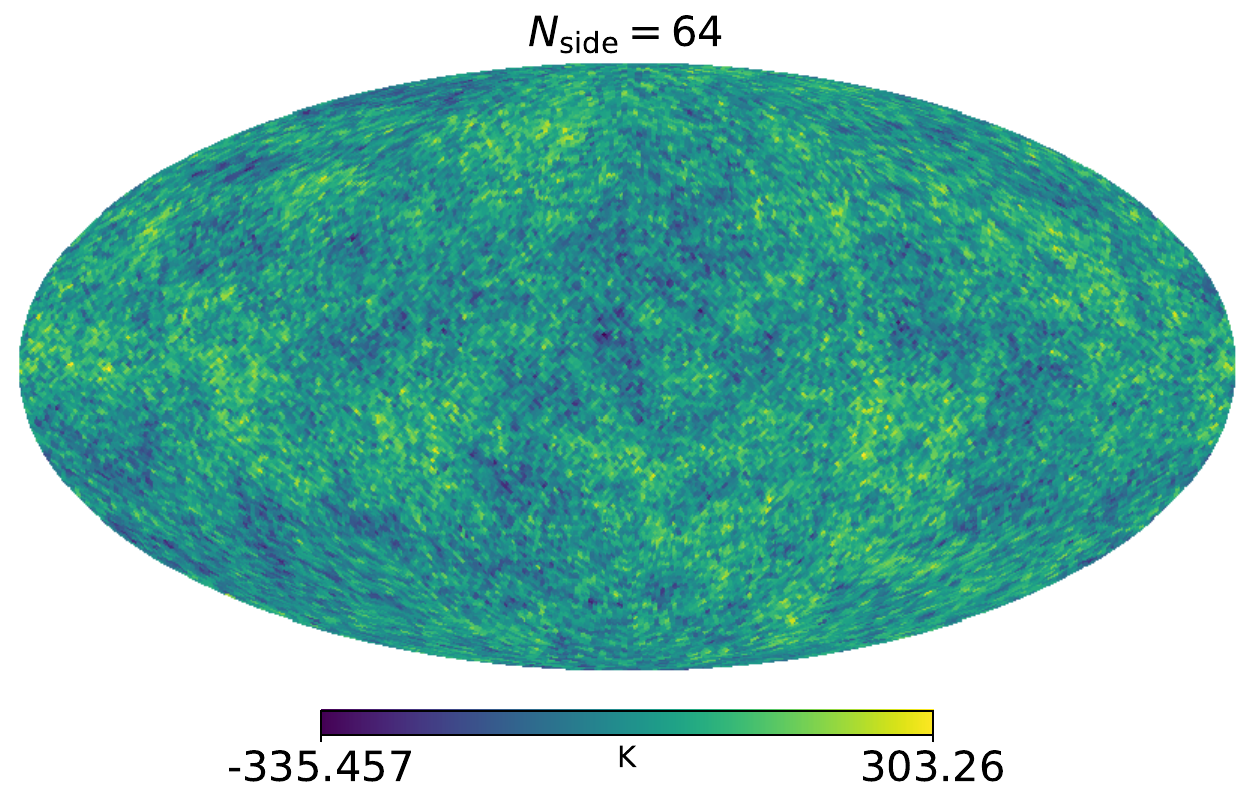}
  \end{minipage}\hfill
  \begin{minipage}[t]{0.48\textwidth}
    \centering
    \includegraphics[height=\panelheight]{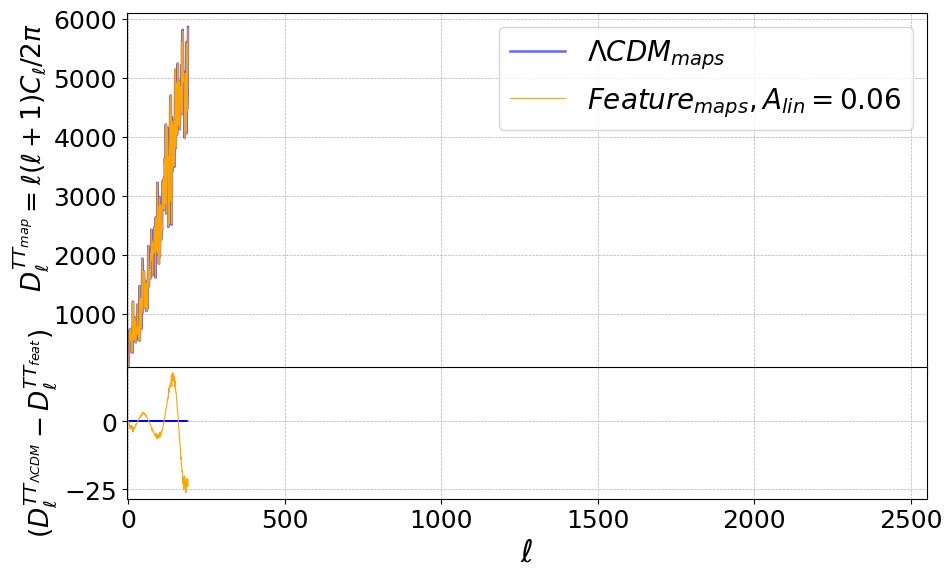}
  \end{minipage}

  \par\vspace{-1mm}
  \begin{minipage}[t]{0.45\textwidth}
    \centering
    \includegraphics[height=\panelheight]{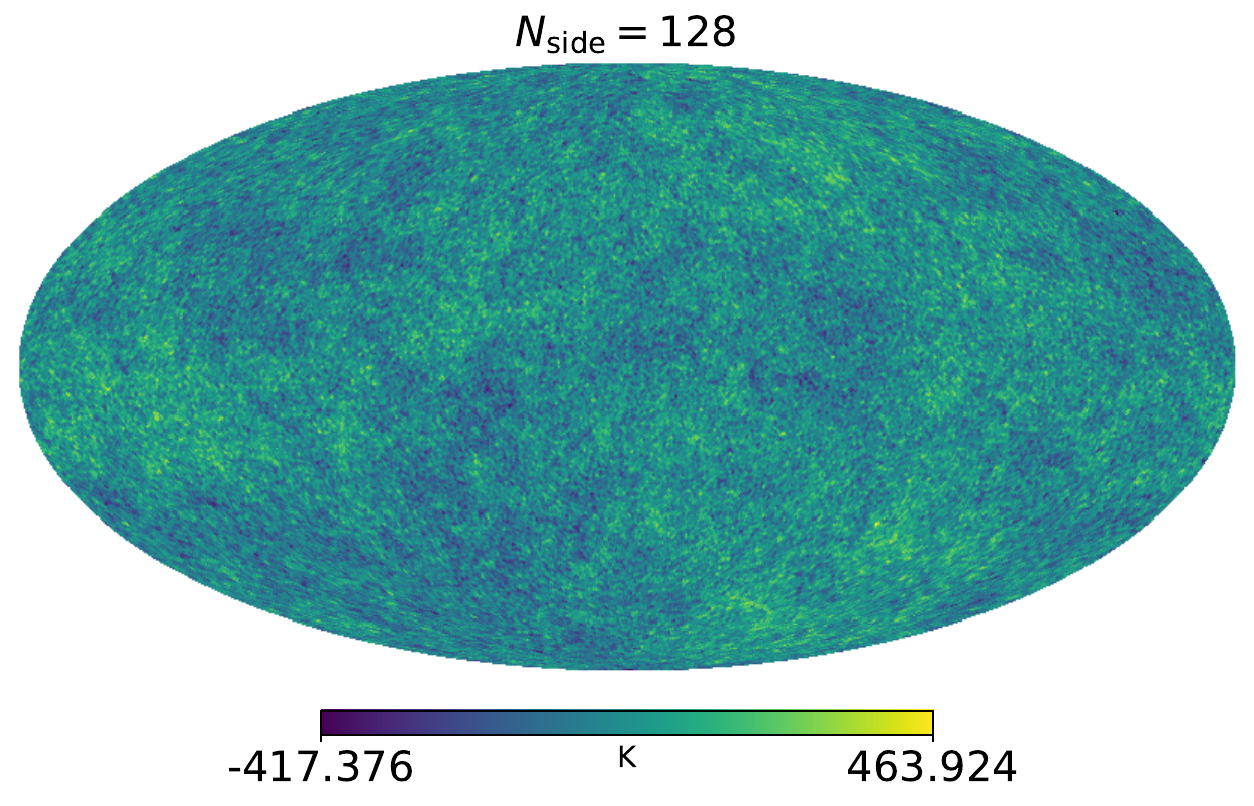}
  \end{minipage}\hfill
  \begin{minipage}[t]{0.48\textwidth}
    \centering
    \includegraphics[height=\panelheight]{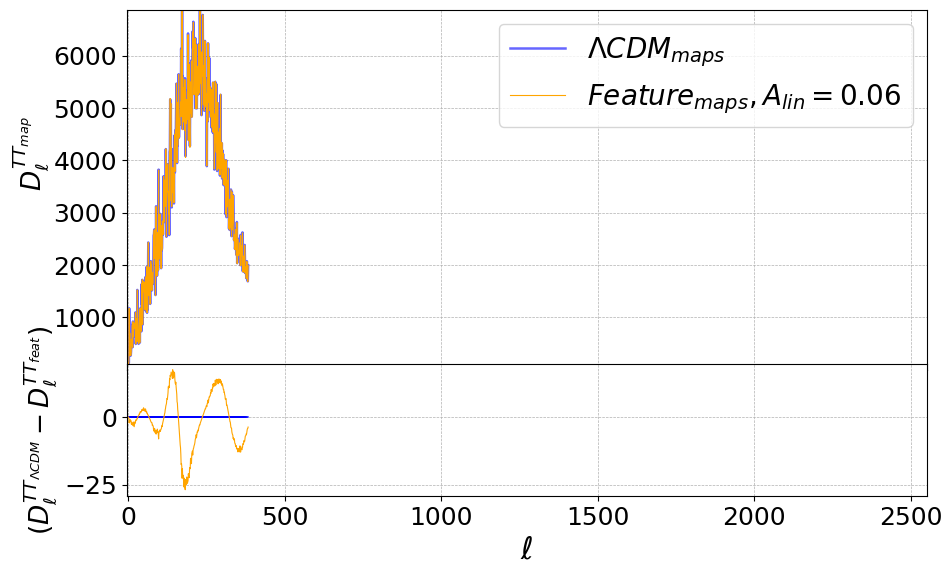}
  \end{minipage}

  \par\vspace{-1mm}
  \begin{minipage}[t]{0.45\textwidth}
    \centering
    \includegraphics[height=\panelheight]{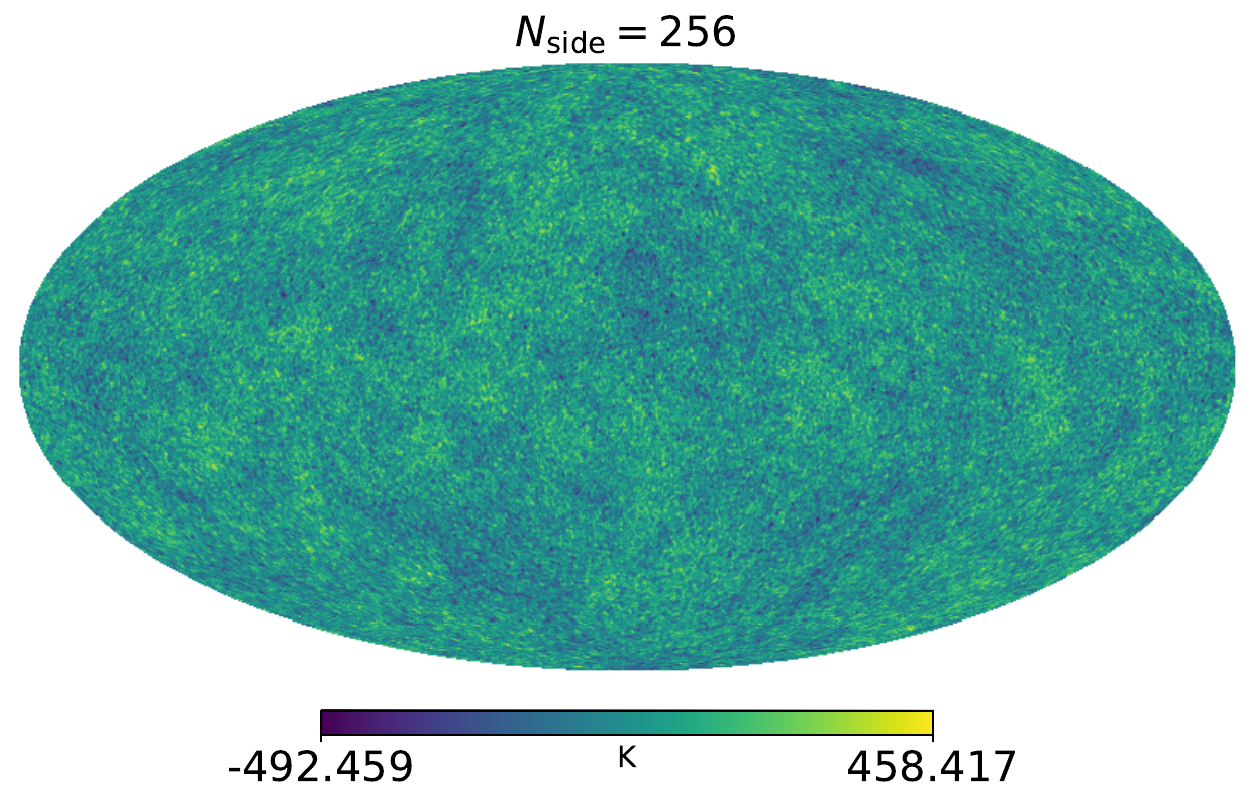}
  \end{minipage}\hfill
  \begin{minipage}[t]{0.48\textwidth}
    \centering
    \includegraphics[height=\panelheight]{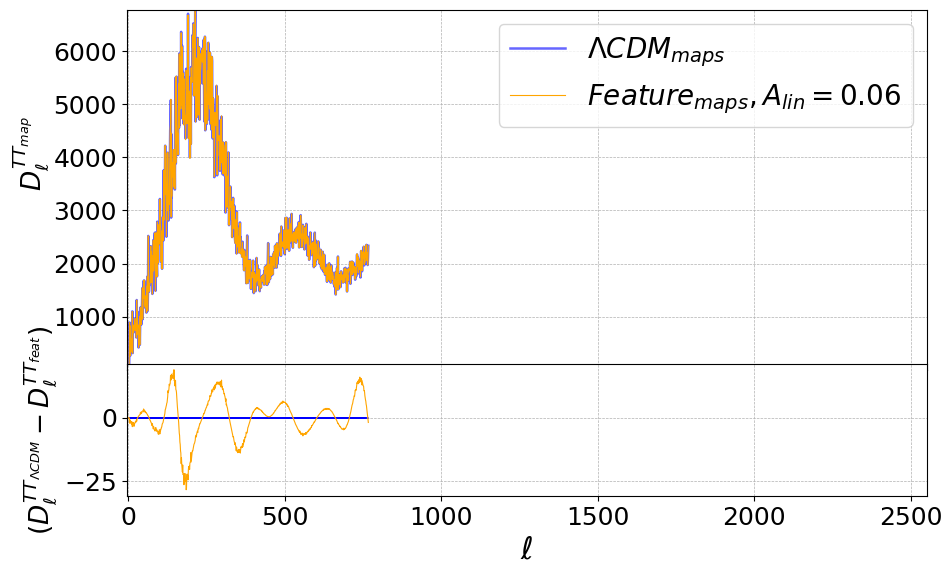}
  \end{minipage}

  \par\vspace{-1mm}
  \begin{minipage}[t]{0.45\textwidth}
    \centering
    \includegraphics[height=\panelheight]{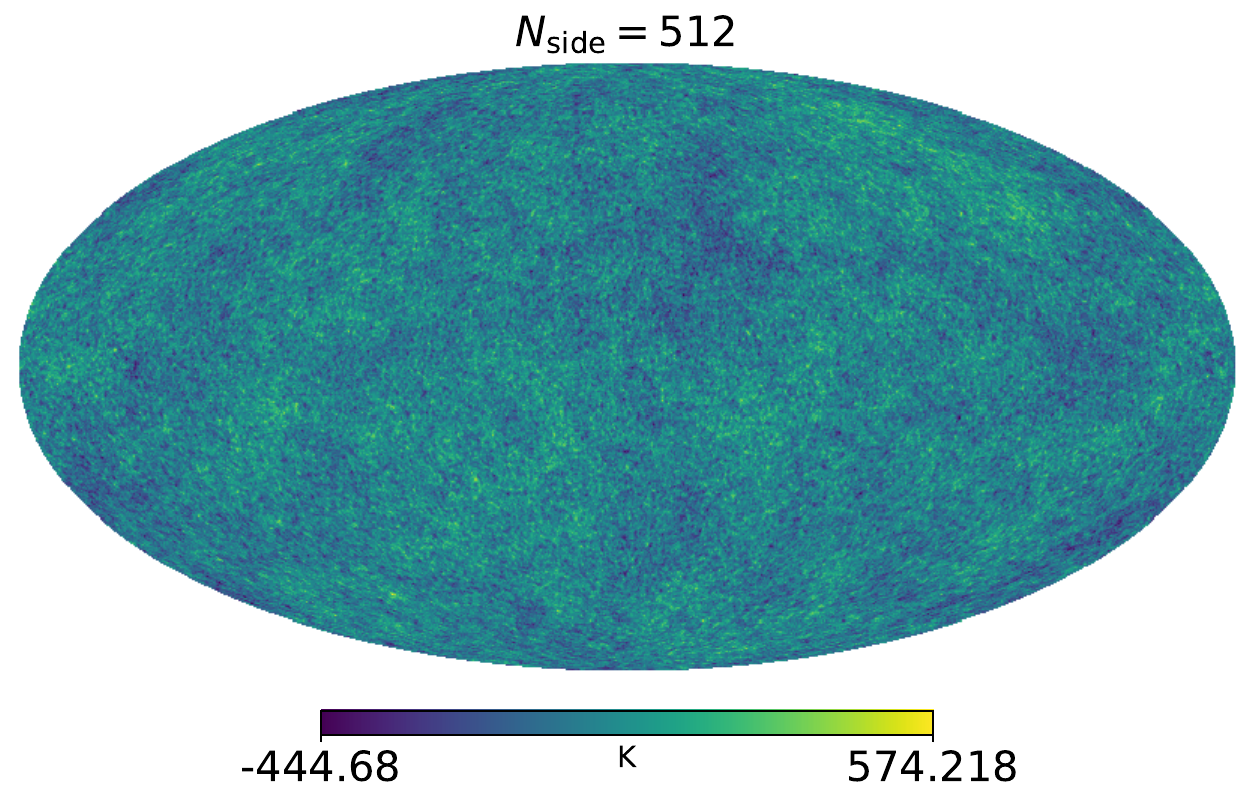}
  \end{minipage}\hfill
  \begin{minipage}[t]{0.48\textwidth}
    \centering
    \includegraphics[height=\panelheight]{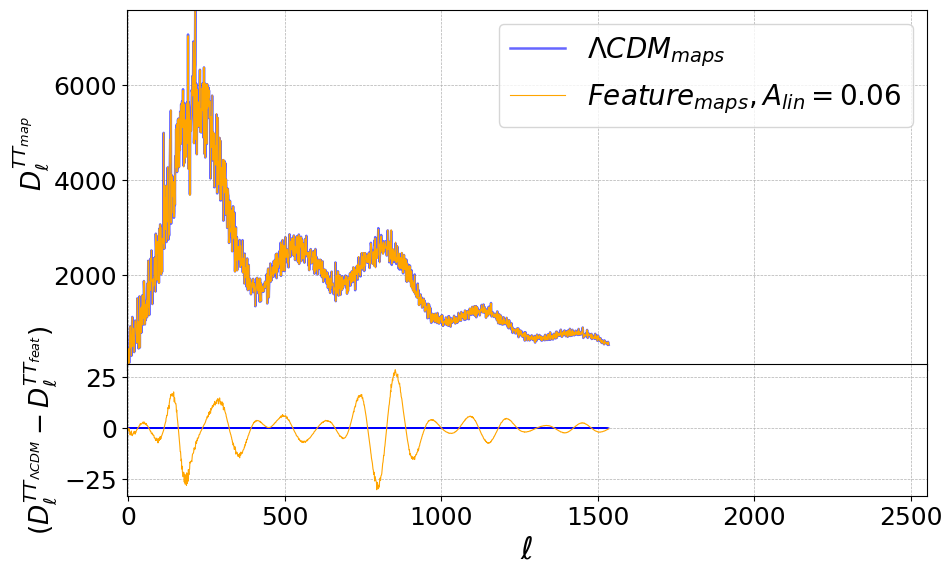}
  \end{minipage}

  \par\vspace{-1mm}
  \begin{minipage}[t]{0.45\textwidth}
    \centering
    \includegraphics[height=\panelheight]{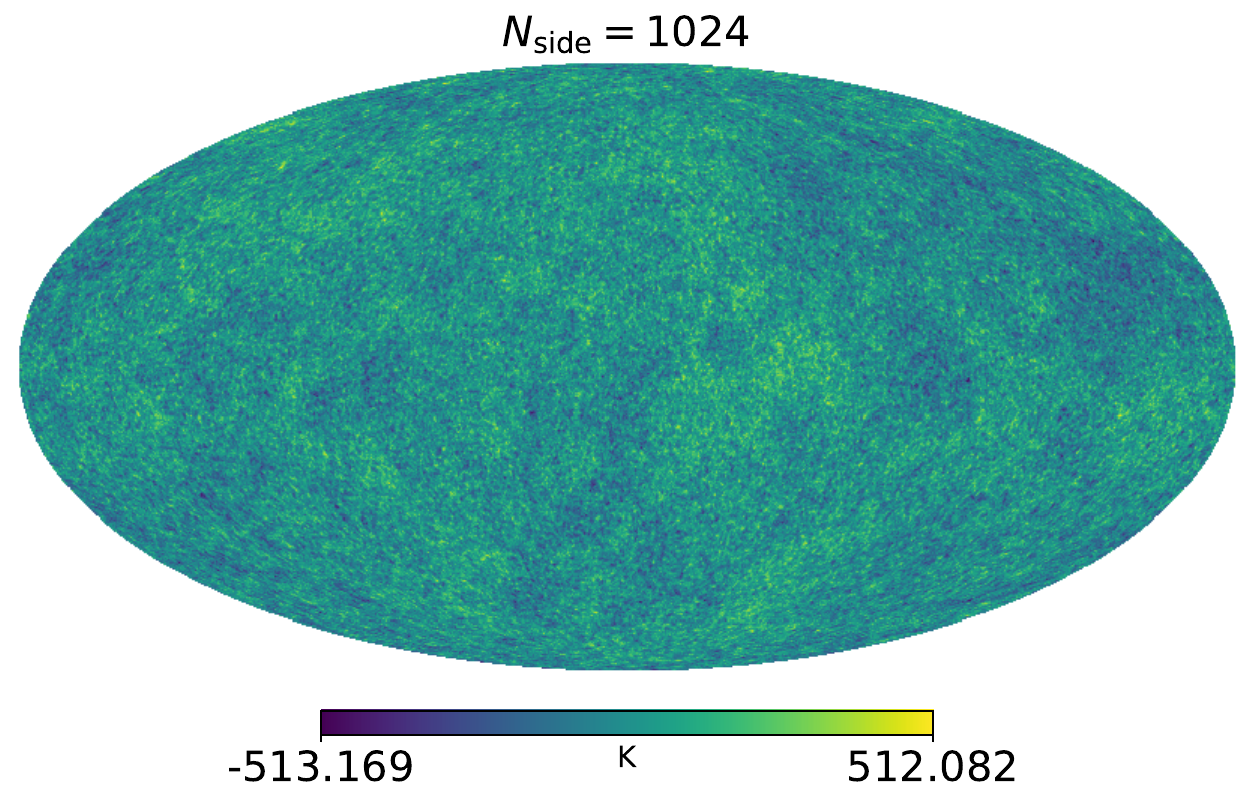}
  \end{minipage}\hfill
  \begin{minipage}[t]{0.48\textwidth}
    \centering
    \includegraphics[height=\panelheight]{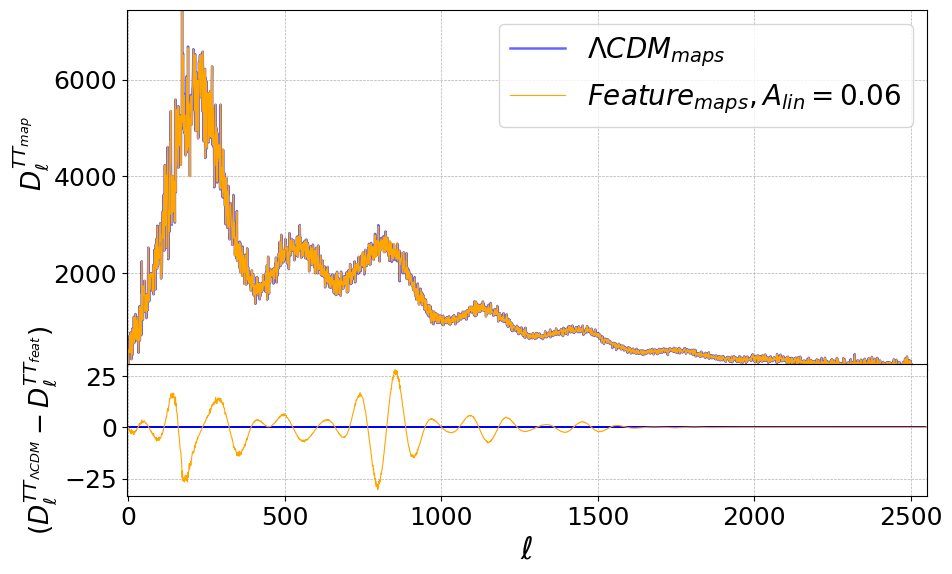}
  \end{minipage}

\caption{Simulated CMB temperature maps (left) and corresponding angular power spectra (right) for \LCDM and a feature model with amplitude $A_{\text{lin}} = 0.06$, generated at different HEALPix resolutions ($N_{\text{side}} = 64, 128, 256, 512, 1024$). The maps illustrate how increasing resolution enhances the small-scale structure of the CMB fluctuations. On the right, the angular power spectra $C_{\ell}$ from the \LCDM baseline (blue) and feature model (orange) are compared, with the lower panels showing the residuals.}
\label{fig:NSIDEvsSignal}
\end{figure}

We adopt a dimensionality reduction strategy combined with a light NN architecture for a balance between computational time and performance. The simulated maps are generated at a HEALPix resolution of $N_{\text{side}}=256$, corresponding to $\sim 8 \times 10^5$ pixels on the sphere \citep{gorski2005healpix}. This choice reflects a compromise between retaining sufficient angular information and ensuring efficiency, as both storage and model complexity scale rapidly with map resolution (see \autoref{app:nn_compilation}). Such resolution is still within the range of the detectable $C_{\ell}$ signal produced by the non-trivial feature considered in the current study. \autoref{fig:NSIDEvsSignal} compares the feature signal recovered at different map resolutions, indicating that $N_{\text{side}}=256$ provides a reasonable balance between retaining the relevant spectral information and limiting computational cost. As illustrated in \autoref{fig:diffmaps}, the maps of the two models, \LCDM and non-standard feature, are visually indistinguishable at the map level. However, the residual map isolates the shift in the global fluctuation power introduced by the feature model. The difference between the two models does not manifest as localised structural features, but as a statistically significant variation in the overall variance of the field. 
The simulated datasets are publicly available on Zenodo\footnote{T, Q, U maps: \url{https://doi.org/10.5281/zenodo.19445834}}.

\subsection{Data pre-processing} \label{subsec:Preprocessing}
Data normalisation aims to transform the values of the dataset into a scale that would enhance the detection of the feature model, due to the high variability of the signals \cite{bhanja2019impact}. Working directly with masked CMB maps, allows the model to retain spatial information that is lost in the compression to $C_{\ell}$, particularly in the presence of a galactic mask which introduces anisotropic sky coverage. We explored several pre-processing techniques to enhance feature detection amongst the high variability of the signals, including wavelet decomposition, high-pass filtering, and standard Z-score normalisation. Although these methods were tested in combination with various Convolutional Neural Network (CNN) architectures, they struggled to consistently isolate the subtle spectral distortions of the feature model from the overwhelming variance of the standard \LCDM signal. The most effective pipeline, and the one adopted for this study, is a per-map standardisation \citep{wang2022recovering}. For the standardisation, each map $X_i$ is transformed as follows.

\begin{equation}
X_i^{\prime}=\frac{X_i-\mu_i}{\sigma_i+\epsilon}
\end{equation}

where $\mu_i$ and $\sigma_i$ represent the mean and standard deviation of that specific pixels of the map. We normalise the maps to make the classification invariant to the global power spectrum amplitude. This forces the model to identify features based on the spatial distribution and shape of the CMB fluctuations. For the classification of the simulated maps, we implemented a hybrid architecture that couples principal component analysis (PCA) with a multilayer perceptron (MLP), which is an artificial NN, consisting of at least three layers: input, hidden, and output. This will be detailed in the next section, which discusses the architecture. 

\subsection{Neural network architecture} \label{subsec:NNarchitecture}
Our classification pipeline is trained on the $(T,\;Q,\;U)$ maps, and thus has access to the information contained in both temperature and linear polarisation. The high dimensionality of the standardised spherical maps, together with the geometry of the galactic mask, presents a challenge for traditional (randomised) weight initialisation in NNs. To mitigate this, we employ a hybrid architecture that leverages PCA as a feature extraction layer, followed by a non-linear MLP part implemented using \texttt{TensorFlow} (version 2.10.1).

The initial stage of the model performs a linear dimensionality reduction. The standardised maps, originally of shape $(H,\;W,\;C)$, are flattened into one-dimensional vectors $x \in \mathbb{R}^D$, where $D$ denotes the number of unmasked pixels. We fit a PCA transformation on the training set and retain the $k = 200$ leading components, which capture the dominant variance of both \LCDM and the feature model.

The PCA is implemented within the \texttt{TensorFlow} framework as a dense layer with weights $W \in \mathbb{R}^{D \times k}$ and bias $b \in \mathbb{R}^k$. The resulting projection is given by
\begin{equation}
z = xW + b.
\end{equation}
This parametrisation is equivalent to the standard PCA transformation
\begin{equation}
z = W^\top (x - \mu),
\end{equation}
where $\mu$ denotes the mean of the training data. In this formulation, the bias term encodes the centering of the data, such that $b = -\mu W$. Implementing PCA as the first learnable layer of a NN is standard in the PCA‑initialised networks \cite{wang2020pca, ren2016convolutional}.

The weights $W$ correspond to the leading eigenvectors of the covariance matrix and are fixed after fitting, which keeps this layer frozen during training. Restricting the input to these principal components provides a regularised representation of the maps. This prevents the network from fitting to stochastic noise in the high-dimensional space, and enhances the model's transferability to different realisations.

Statistically, PCA reorients the data along a new set of independent axes, or eigenvectors, derived from the covariance matrix $C = \langle x x^\top \rangle$. The primary components, those with the highest variance, correspond to a condensed representation of the maps that preserves the most significant features while discarding redundant information \citep{efstathiou2002principal}. In the context of CMB maps, this allows the model to focus on the dominant statistical structure of the field while significantly reducing the dimensionality of the input space.

The resulting weights ${W}$ and biases ${b}$ were embedded as a fixed, non-trainable linear layer at the input of the NN, designed to capture non-linear relationships between the principal components. In contrast to linear models, NNs can learn complex mappings between inputs and outputs by combining simple operations across multiple layers.

Since we train on T, U and Q maps, we implement 3 similar architectures detailed in \autoref{tab:architecture}. For the three of them, the first stage applies a non-linear activation function, specifically the Rectified Linear Unit (ReLU), defined as $\mathrm{ReLU}(x) = \max(0, x)$, which introduces non-linearity and enables the network to model complex feature interactions \citep{nair2010relu}. The use of ReLU activations in this layer allows the model to learn non-linear combinations of the $k$ principal components. This is followed by a fully connected (dense) hidden layer with $n$ neurons (see \autoref{tab:architecture}), each neuron forms a weighted combination of all input data. After this, we include a dropout layer, with a dropout rate specified in \autoref{tab:architecture}.

\begin{table}[h!]
\centering
\begin{tabular}{|c|c|c|c|c|}
\hline
\textbf{Maps} & \textbf{Activation} & \textbf{Hidden layer (n)} & \textbf{Dropout rate} & \textbf{Output activation} \\
\hline
\hline
T & ReLU & 64 & 0 & No\\
U & ReLU & 128 & 0.2 & No\\
Q & ReLU & 128 & 0.2 & No\\
\hline
\end{tabular}
\caption{Hyperparameters for the MLP classification part integrated into the hybrid PCA-NN architecture. Separate configurations are shown for the Temperature ($T$) and polarisation ($U, Q$) map pipelines, with $n$ number of layers. The Output activation column indicates that raw logits are produced to optimise the binary cross-entropy loss calculation.}
\label{tab:architecture}
\end{table}

The output layer consists of a single neuron with a linear activation, producing a real-valued score (referred to as logit), which is subsequently transformed into a probability through a sigmoid function \citep{goodfellow2016deep}. This formulation is standard for binary classification problems. The model parameters are optimised using the Adam algorithm, a stochastic gradient-based optimiser that adapts the learning rate for each parameter during training \citep{kingma2014adam}. The loss function is Binary Cross-Entropy, to quantify the discrepancy between predicted probabilities and true class labels, widely used for probabilistic binary classification tasks \citep{goodfellow2016deep}.

To mitigate overfitting and ensure robust generalisation, we employ the Early Stopping callback based on the validation performance. Specifically, we monitor the Area Under the Receiver Operating Characteristic Curve (AUC), a threshold-independent metric that measures the model’s ability to discriminate between classes \citep{fawcett2006roc}. Training is stopped if the validation AUC does not improve for 50 consecutive epochs, and the model weights corresponding to the best performance are restored.

This architecture allows the model to learn subtle, non-linear correlations between modes that are not captured by standard summary statistics. The projection of the maps onto the leading PCA components, helps obtain a compressed representation that captures the dominant variance of the data.

In the context of CMB maps, the principal components encode the dominant covariance structure of the field, closely related to the angular power spectrum. This provides a compact representation of the main cosmological information, allowing the architecture to efficiently evaluate the shifts in global variance introduced by the non-standard feature model directly from the pixel-level data.
The trained NN weights and the model are publicly available on Zenodo\footnote{NN model weights: \url{https://doi.org/10.5281/zenodo.19445671}}.

\subsection{Validation of the learning pipeline and robustness tests}

To assess the robustness of our classification framework, we performed a series of validation tests on the data pre-processing and training pipeline. First, we verified that the PCA transformation was fitted exclusively on the training data, which prevents from information leakage into the validation and test sets. We also confirmed that the statistical properties (mean, variance, and dynamic range) of the input maps were consistent across all data splits.

We conducted several diagnostic tests to evaluate how effectively different pipelines can extract the cosmological signal directly from pixel-space data. Convolutional and fully connected NN were trained directly on raw and standardised maps, but were unable to overfit even small subsets of the training data. This behaviour suggests that the discriminative signal between \LCDM and the feature model is weak in pixel space and is dominated by global statistical variations.

We then investigated whether the limited performance of CNNs could be attributed to optimisation difficulties. CNNs trained with first-order gradient-based optimisers, such as Adam and stochastic gradient descent, failed to converge to meaningful solutions. In contrast, linear models trained using convex optimisation procedures achieved stable and reproducible performance. This difference is consistent with the fact that linear classifiers correspond to convex optimisation problems with a unique global solution, whereas NNs involve highly non-convex loss that can be difficult to optimise in high-dimensional, correlated input spaces~\citep{goodfellow2016deep}.

These results indicate that, although discriminative information is present in the raw maps, the associated optimisation problem is poorly conditioned in pixel space. In this context, PCA is introduced as an explicit first stage of the model, rather than as a pre-processing step. By projecting the data onto an orthogonal basis of leading variance directions, PCA reduces feature correlations and concentrates the relevant information into a lower-dimensional subspace. This transformation improves the conditioning of the learning problem and enables stable training with gradient-based methods.

Therefore, we combined PCA with a MLP because it provides a more effective representation for classification. In particular, the PCA stage suppresses dominant \LCDM variance and enhances residual structures associated with the primordial feature, making it more accessible to the next non-linear classification layers, and it is particularly suited for the final interpretability analyses based on \texttt{SHAP}.

\subsection{Post-hoc interpretability}
We apply \texttt{SHAP} for the interpretability analysis of our architecture's predictions\footnote{\url{http://github.com/shap/shap}}. This framework assigns each input pixel $i$ an attribution value, $\phi_i$, representing its contribution to the model's output $f(x)$. Mathematically, these values are derived from the Shapley values in cooperative game theory, defined as the weighted average of marginal contributions across all possible input subsets $S \subseteq M \setminus \{i\}$:

\begin{equation}
\phi_i=\sum_{S \subseteq M \backslash\{i\}} \frac{|S|!(M-|S|-1)!}{M!}\left[f_S(x \cup\{i\})-f_S(x)\right]
\end{equation}

where $M$ is the total number of features\footnote{Referring to the input data. Not to be confused with the non-standard feature model} (pixels) and $f_S$ is the model's prediction restricted to the subset $S$. The term $\left[f_S(x \cup\{i\})-f_S(x)\right]$ shows how the prediction changes when pixel $i$ is added to the knowledge base.

\begin{figure}[h!]
\centering
\includegraphics[width=1.01\linewidth]{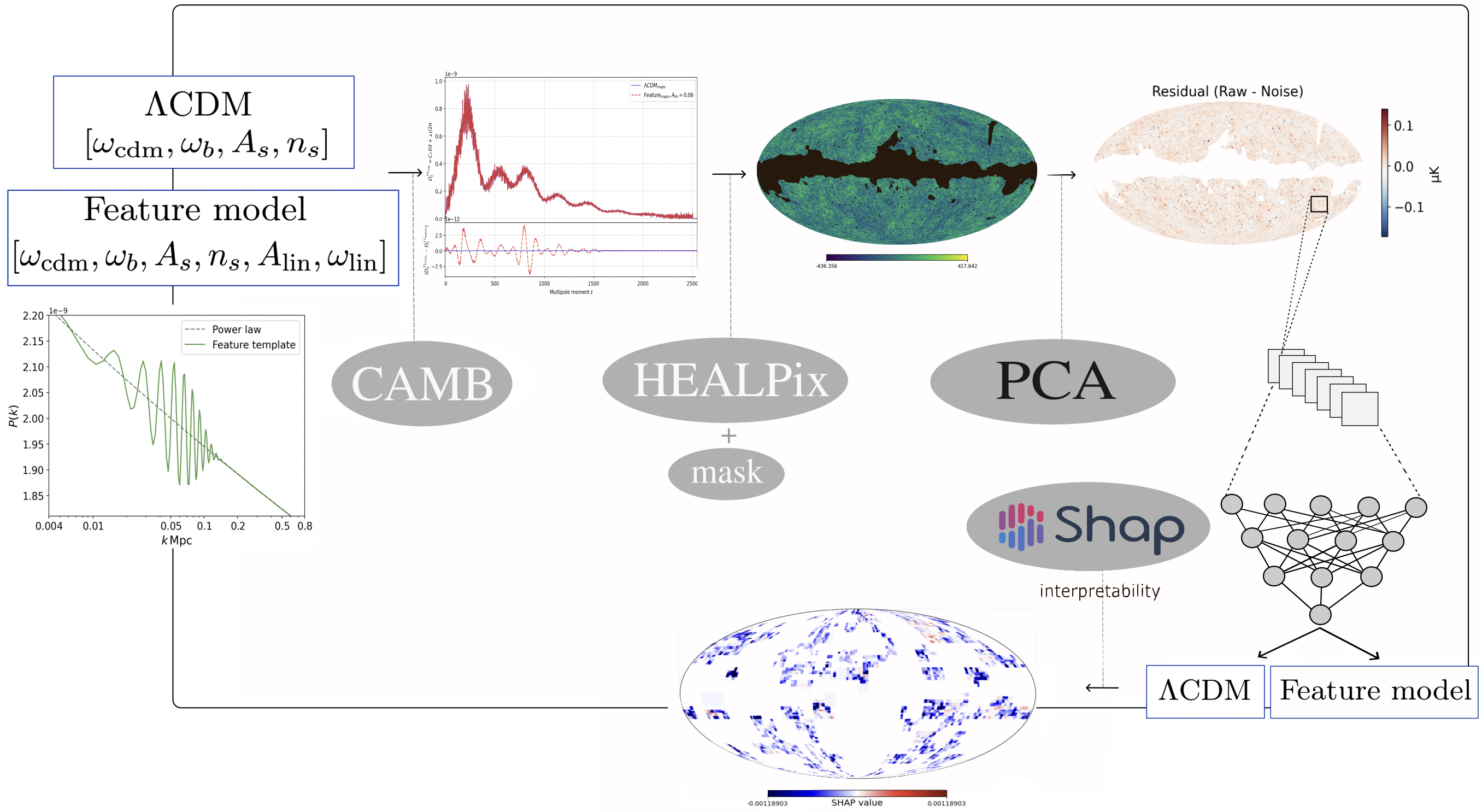}
\caption{Overview of the end-to-end simulation, classification and interpretability pipeline for the temperature CMB maps. Starting from cosmological parameters for both \LCDM and the non-standard feature model, we use CAMB and a Planck-like covariance matrix to generate primordial power spectra. These are transformed into full-sky maps via HEALPix, where a galactic mask is applied. A PCA-based residual analysis is used to isolate relevant features before the maps are passed to a NN for model classification. Finally, the \texttt{SHAP} framework is applied to produce map-level interpretability visualisations, identifying the spatial regions that contribute most to the architecture's decision-making process.} \label{fig:scheme}
\end{figure} 

To calculate the pixel attribution, we employed the \texttt{Partition} SHAP algorithm \cite{lundberg2020local}. Unlike standard sampling approaches, \texttt{Partition} SHAP employs a hierarchical clustering of the input pixels (using a `masker') to account for spatial correlations. It computes Shapley values $\phi_i$ by recursively partitioning the image into a tree structure and evaluating the marginal contribution of these groups to the model’s logit output. This approach is well-suited for HEALPix maps, as it maintains computational efficiency while also ensures that the resulting importance values reflect the signal of the primordial feature model. By projecting these $\phi$ values back onto the HEALPix sphere, we generate spatial heatmaps of importance. These visualisations allow us to assess whether the network's evaluation of the global variance, which is introduced by the $A_{\mathrm{lin}}$ and $\omega_{\mathrm{lin}}$ parameters, is properly distributed across the statistically isotropic unmasked sky. 
This step is important for providing evidence that the classification observed in the previous section is consisted with the feature template that we used to simulate half of the CMB maps, providing a transparent link between the high-dimensional map space and some parameters.

The integrated simulation and classification workflow is summarised in~\autoref{fig:scheme}, tracing the pipeline from primordial power spectra generation via CAMB to the production of masked HEALPix maps. We employ a PCA-based dimensionality reduction to extract the dominant modes of variance, which serve as the input for a NN classifier. To validate the model's decision-making process, we apply the \texttt{SHAP} framework to map the network's internal logic back onto the spatial sky. This approach supports the interpretation that the discrimination between \LCDM and the feature model is driven by an overall fluctuation power of the cosmological signal. The attribution maps generated with \texttt{SHAP} are availbale on Zenodo.\footnote{NN interpretability attribution maps: \url{https://doi.org/10.5281/zenodo.19445676}}\\

\section{Results}\label{sec:results}

In this section, we present the performance of the classification framework introduced in \autoref{sec:methodology}. Our analysis focuses on assessing the ability of the model to distinguish between \LCDM and the non-standard feature model using simulated CMB maps. More importantly, the results of the interpretability part using \texttt{SHAP}.

\subsection{Classification of models}\label{subsec:classification}
\begin{figure}[ht]
  \raggedright 
  \newcommand{\panelheight}{0.16\textheight}
  
  \begin{minipage}[t]{0.305\textwidth}
    \includegraphics[height=\panelheight]{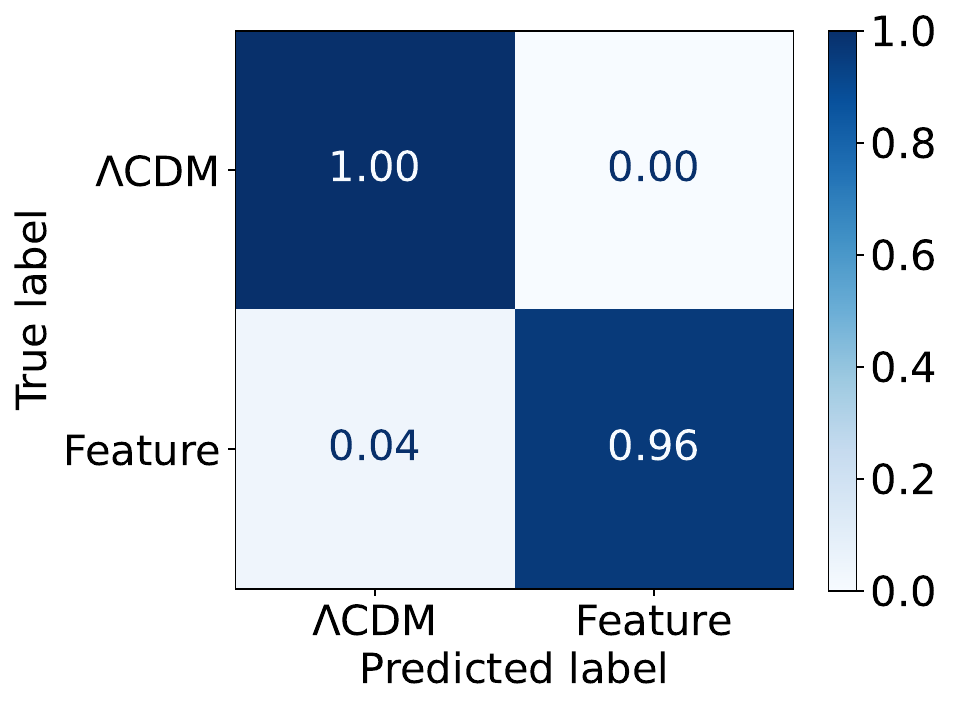}
  \end{minipage}
    \hspace{1em}
  \begin{minipage}[t]{0.305\textwidth}
    \includegraphics[height=\panelheight]{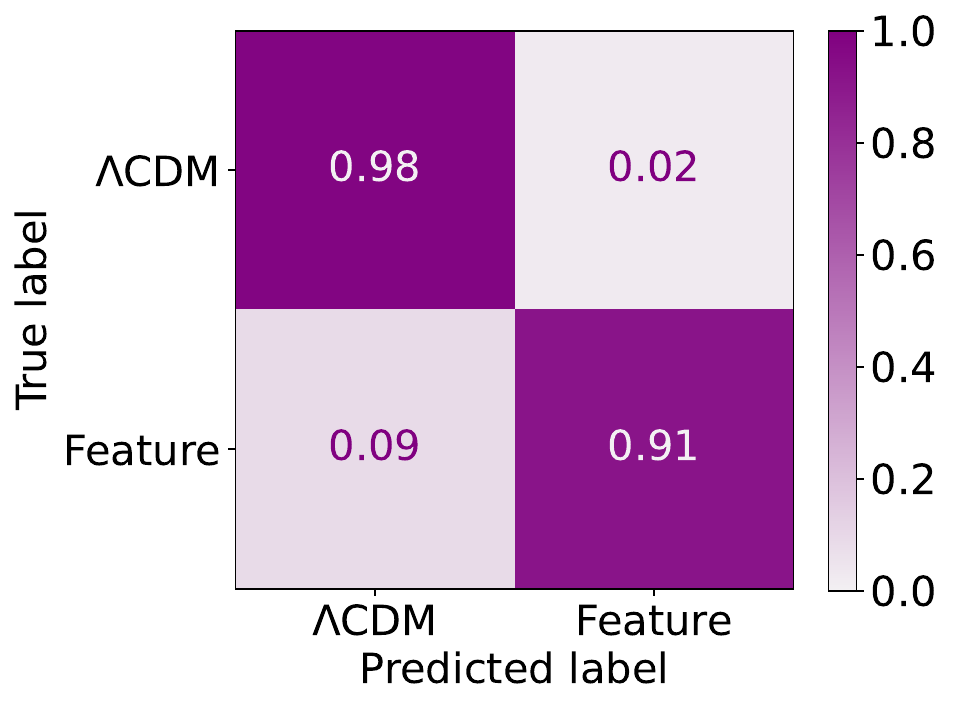}
  \end{minipage}
  \hspace{1em}
  \begin{minipage}[t]{0.32\textwidth}
    \includegraphics[height=\panelheight]{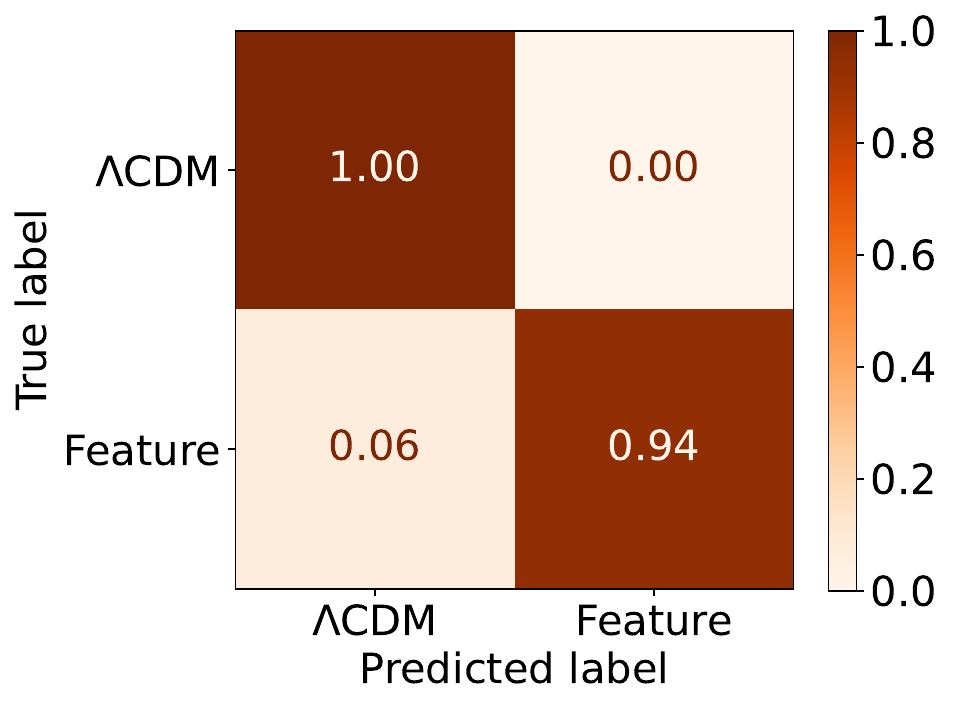}
  \end{minipage}

  \caption{Confusion matrices reflecting the performance of the CNNs for a feature of amplitude $A_{\text{lin}}\in[0.01, 0.06]$ and $\omega_{\text{lin}}\in[5, 25]$. Left: T maps, center: U maps, right: Q maps.}
  \label{fig:ConfMatrix}
\end{figure}

\autoref{fig:ConfMatrix} shows the confusion matrices obtained for the classification of \LCDM and the feature model with feature amplitude and frequency ranges indicated in~\autoref{tab:prior} using the PCA-preprocessed maps and the MLP. The signal, as we could notice in \autoref{fig:diffmaps}, is imprinted globally in the temperature $T$ and polarisation $U,\;Q$ maps. In this regime, the classifier achieves a really good discrimination between the two classes on the test set, within the considered parameter range.

This high accuracy is achieved thanks to the PCA-based dimensionality reduction, which projects the data onto the dominant modes of variation of the training set. This transformation provides a compact and structured representation of the maps, filtering out redundancy in the high-dimensional pixel space and highlighting the most informative directions for classification. As a result, subtle deviations from the standard \LCDM model become more accessible to the NN. When the PCA step is removed, the classification performance drops to near-random levels, indicating that the discriminative signal is difficult to extract directly from raw pixel space.

The performance observed for $A_{\mathrm{lin}} \in [0.01, 0.06]$ and $\omega_{\mathrm{lin}} \in [5, 25]$ therefore reflects the effectiveness of the PCA + MLP pipeline in isolating feature signatures. This configuration provides a controlled setting for subsequent interpretability analyses, allowing us to identify the specific spatial patterns responsible for the model’s decisions using SHAP values.

\subsection{Interpretability of the trained pipeline}\label{subsec:Interpretability}
\begin{figure}[h]
\centering
\includegraphics[width=1.01\textwidth]{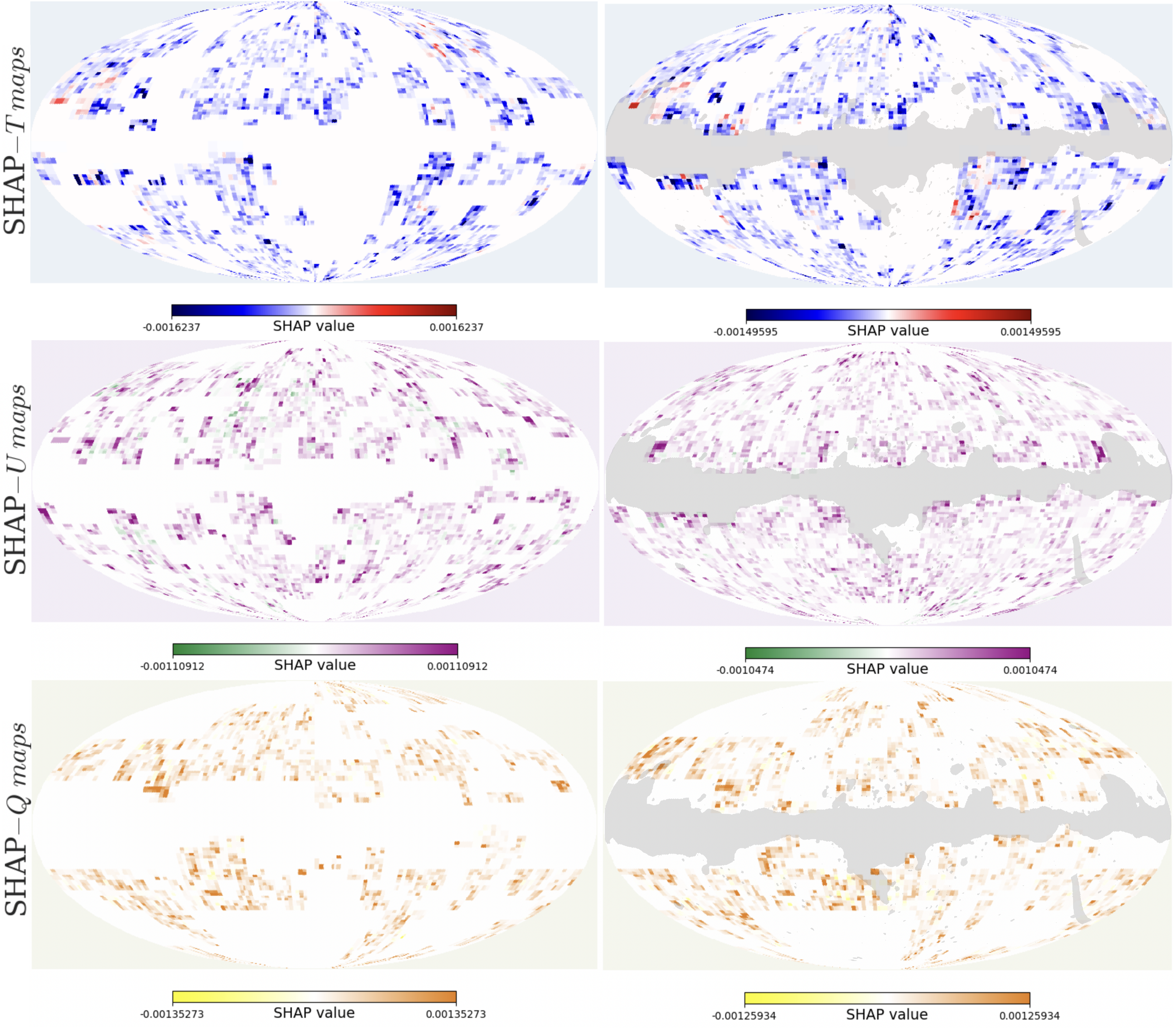}
\caption{\texttt{SHAP} attribution maps for the classification of \LCDM vs primordial feature model. The left column shows the attribution maps obtained with SHAP \cite{lundberg2020local} corresponding to the CMB simulated masked maps. The right column includes the corresponding galactic mask to highlight that these regions are not considered by the network. From top to bottom: Temperature ($T$), Stokes $U$, and Stokes $Q$. These attribution maps correspond only to the primordial feature realisations, which confirms the network responds exclusively to the signatures of this alternative model.}
\label{fig:ShapMaps}
\end{figure}

The spatial distribution of the feature-driving information is visualised via the \texttt{SHAP} attribution maps in \autoref{fig:ShapMaps}. These represent the pixel-wise contribution to the network’s final classification for the Temperature $T$, and Stokes $Q$ and $U$ polarisation components. We observe that the importance values are distributed globally across the high-latitude regions, rather than being concentrated in isolated local places. This suggests that the NN has learned to identify the diffuse, large-scale oscillatory signatures characteristic of the $A_{\mathrm{lin}}$ and $\omega_{\mathrm{lin}}$ feature model employed.

Notably, the SHAP values remain consistent across the masked simulations. The attribution patterns in the masked maps do not show significant border effects near the galactic cut, which confirms that the PCA-NN pipeline successfully prioritises cosmological residuals over masking geometry. The balanced contribution across $T$, $Q$, and $U$ further indicates that the model uses the cross-correlation between temperature and polarisation to reach its classification decision, reinforcing the physical robustness of the architecture.

The discriminative power of the pipeline is further evidenced by the contrast between the attribution maps for the two classes. While the feature model exhibit structured, globally distributed SHAP values across the $T$, $Q$, and $U$ components, the corresponding maps for the baseline \LCDM simulations remain essentially featureless. This response in the \LCDM case indicates that the NN does not find significant evidence for the $A_{\mathrm{lin}}$ and $\omega_{\mathrm{lin}}$ signature in the standard cosmological model. As a result, the high-importance regions in the SHAP maps (\autoref{fig:ShapMaps}) seem to track slightly the signal imprinted by the primordial feature oscillations, see \autoref{fig:diffmaps}. This confirms that the network is identifying the signal itself.

We observe that the SHAP attributions are distributed globally across the unmasked map, consistently reflecting the statistical weight associated with the targeted scales of the primordial features. Since the \LCDM baseline shows no such structured response, we can conclude the architecture is capturing the intended physical difference in variance. The pipeline provides a robust framework for detecting these non-standard cosmological signatures by distinguishing the modified global variance from the baseline \LCDM statistics, while remaining unaffected by the geometric constraints of the galactic mask. These results suggest that the integrated PCA-NN approach is a reliable tool for identifying non-standard features explicitly within the primordial power spectrum. We emphasise that SHAP provides a diagnostic of model behaviour and decision-making, rather than a proof of localized physical causation.\\

\section{Conclusions}\label{sec:conclusions}
In this work, we have presented a robust ML framework designed to discriminate between the standard \LCDM model and an alternative scenario featuring primordial power spectrum oscillations. We successfully reduced the high-dimensional complexity of masked Planck-like temperature and polarisation maps while preserving the statistical structure of the data, by integrating PCA as a fixed, non-trainable input layer within a NN architecture. Our results demonstrate that this hybrid PCA–NN pipeline achieves really good classification accuracy at feature amplitudes of $A_{\mathrm{lin}} \in [0.01, 0.06]$ and frequencies of $\omega_{\mathrm{lin}} \in [5, 25]$, even in the presence of galactic masking and realistic instrumental noise. We find that the PCA stage, which projects the data onto the leading variance modes of the training set, is essential for feature recovery; without this structured dimensionality reduction, the discriminative signal seems inaccessible in the raw pixel space.

We also implemented the \texttt{SHAP} interpretability framework, which serves as a diagnostic part of the model’s performance. By decomposing the classification logic into pixel contributions, the resulting attribution maps demonstrate that the network’s decision-making process is driven by globally distributed signatures across the Temperature ($T$) and Stokes ($Q, U$) polarisation planes. The widespread distribution of these features suggests that the network is sensitive to the oscillatory residuals present throughout the map. This indicates that the model is learning the actual signal rather than being biased by localised instrumental noise or the edges of the galactic mask. The lack of signal in the \LCDM attribution maps, acts as a control, demonstrating that the NNs detections are specific to the primordial features. Therefore, we can conclude that the classification achieved is a direct consequence of the model successfully isolating the signal produced by the non-standard feature template in the primordial power spectrum.

The decision to perform the analysis directly on the sky maps, rather than on the estimated angular power spectra $C_\ell$, is driven by the aim of developing a diagnostic pipeline for CMB maps. Moreover, we included a galactic mask in the simulated maps, that served as a benchmark to corroborate that the architecture is extracting the information from the unmasked sky. While a galactic mask inevitably introduces mode coupling that limits the available cosmological information, our PCA-NN framework processes the pixel data directly, avoiding the need for explicit harmonic deconvolution steps. This allows us to verify that the network derives its statistical evaluations from the unmasked high-latitude regions while remaining robust to the discontinuous boundaries of the galactic cut. Consequently, the network can assess the changes in overall fluctuation power associated with the feature model without relying on the explicit harmonic deconvolution required by masked power-spectrum estimators.


This end-to-end pipeline offers a scalable and interpretable alternative to traditional likelihood based searches for primordial features. While this study focused on linear oscillations in Fourier space, the flexibility of the PCA-NN architecture allows for future extensions into more complex localised features or non-Gaussian signatures. As the next generation of CMB experiments provides increasingly high-resolution polarisation data, such integrated deep learning tools will be crucial for exploring the physics of the very early universe.

Ultimately, the application of \texttt{SHAP} to the pixel-level CMB maps, allows us to gain some intuition about the most relevant regions that influence our architecture's decisions, establishing a new standard for transparency in ML-based cosmological analyses.\\ 

\section*{Acknowledgments}
The authors would like to thank S. Nesseris and C. Scóccola for useful discussions. IO is supported by the fellowship LCF/BQ/DI22/11940033 from ``la Caixa” Foundation (ID 100010434), the research project PID2021-123012NB-C43 and the Spanish Research Agency (Agencia Estatal de Investigaci\'on) through the Grant IFT Centro de Excelencia Severo Ochoa No CEX2020-001007-S, funded by MCIN/AEI/10.13039/501100011033 and (partly) by the Spanish Research Agency’s Consolidacion Investigadora 2024 grant CNS2024-154430). The research presented in this publication falls within the research line (Astro/Cosmo). IO also thanks ESTEC/ESA for the warm hospitality during the execution of the first part of this project,
and for support from the \href{https://www.cosmos.esa.int/web/esdc/visitor-programme}{ESA Archival Research Visitor Programme}. GCH acknowledges that this project is part of the project UNICORN with file number VI.Veni.242.110 of the research programme Talent Programme Veni Science domain 2024 which is (partly) financed by the Dutch Research Council (NWO) under the grant \url{https://doi.org/10.61686/ZCPQI32997}. The authors acknowledge the use of the Finis Terrae III supercomputer, which is part of the Centro de Supercomputacion de Galicia (CESGA) and is funded by the Ministry of Science and Innovation, Xunta de Galicia and ERDF (European Regional Development Fund).

\bibliographystyle{apsrev4-1}

\clearpage
\bibliography{references}

\clearpage
\begin{appendix}


\section{A: NN Compilation}  
\label{app:nn_compilation}

\begin{figure}[h!]
    \centering
    \includegraphics[width=0.56\linewidth]{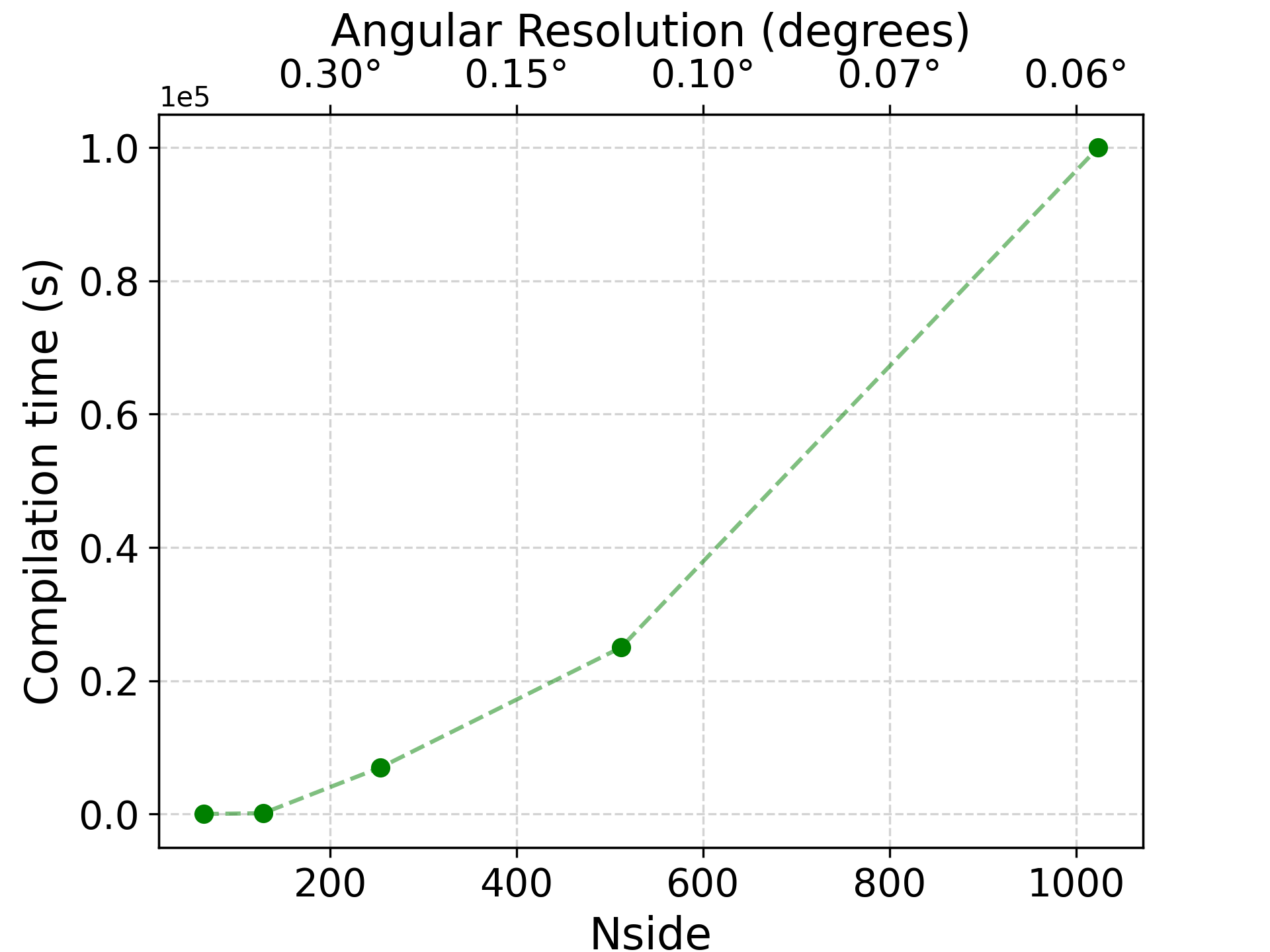}
    \caption{Compilation time of the NN as a function of the HEALPix $N_\text{side}$ parameter.}
    \label{fig:scalability}
\end{figure}

In \autoref{fig:scalability} we can see how the NN compilation time scales with the resolution of the maps. Considering that the feature introduced in the primordial power spectrum is mainly contained up to $N_\text{side} = 256$, we choose this to be the optimal resolution to work with.

\section{B: Receiver Operating Characteristic curves}
\label{app:ROC}
\begin{figure}[htbp]
    \centering
    
    \begin{minipage}{0.32\textwidth}
        \centering
        \includegraphics[width=\linewidth]{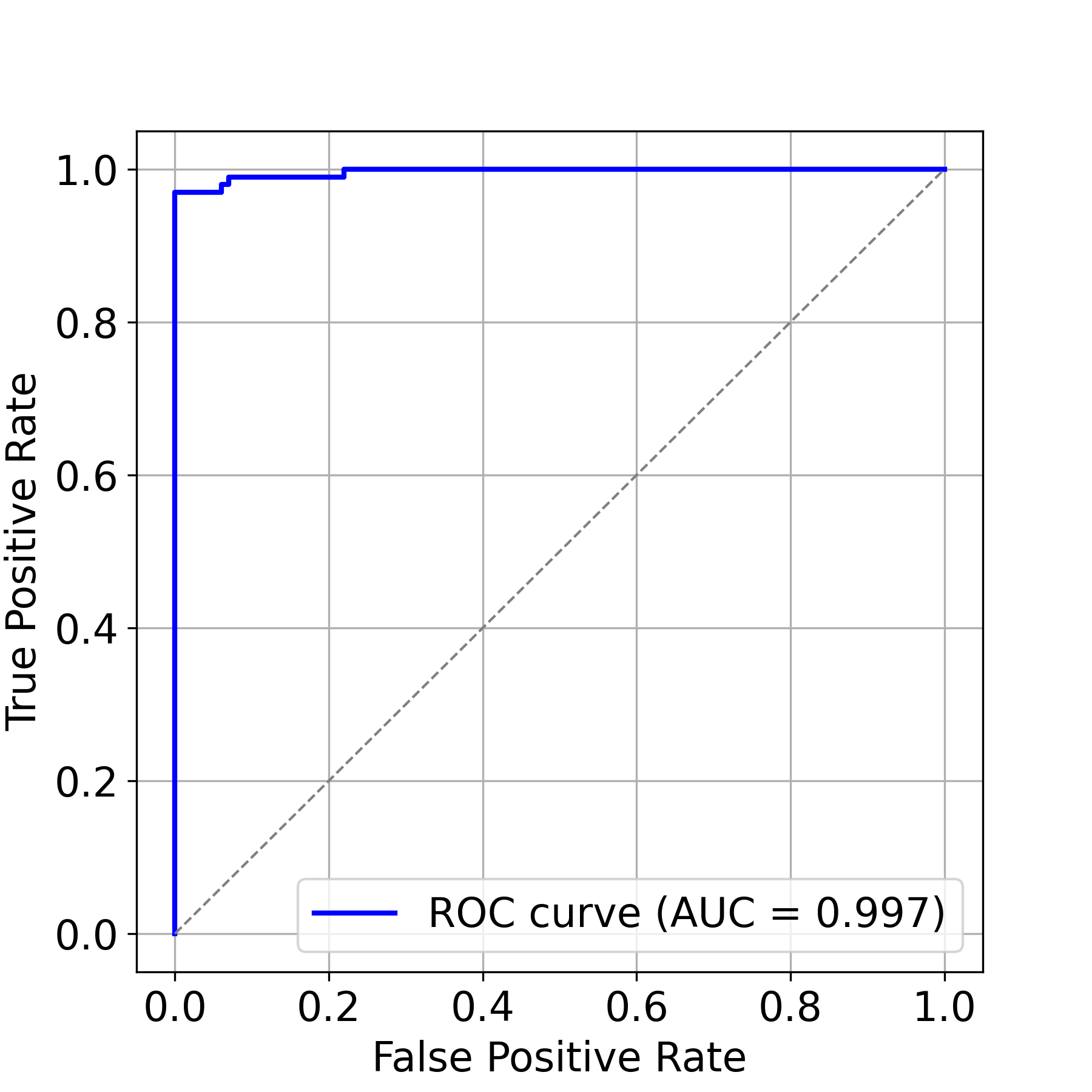}
        \vspace{0.1cm} 
        (a)
    \end{minipage}\hfill
    \begin{minipage}{0.32\textwidth}
        \centering
        \includegraphics[width=\linewidth]{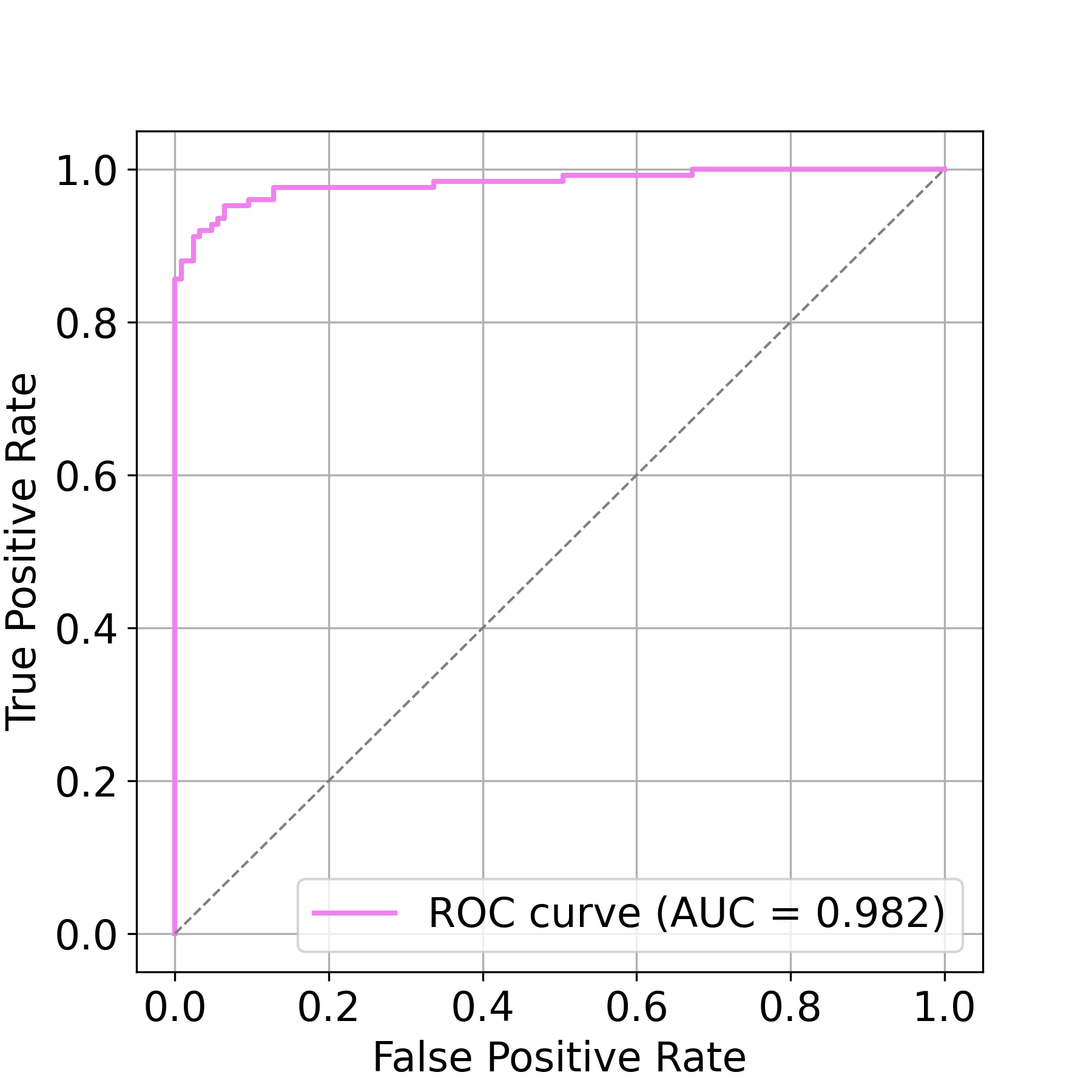}
        \vspace{0.1cm}
        (b)
    \end{minipage}\hfill
    \begin{minipage}{0.32\textwidth}
        \centering
        \includegraphics[width=\linewidth]{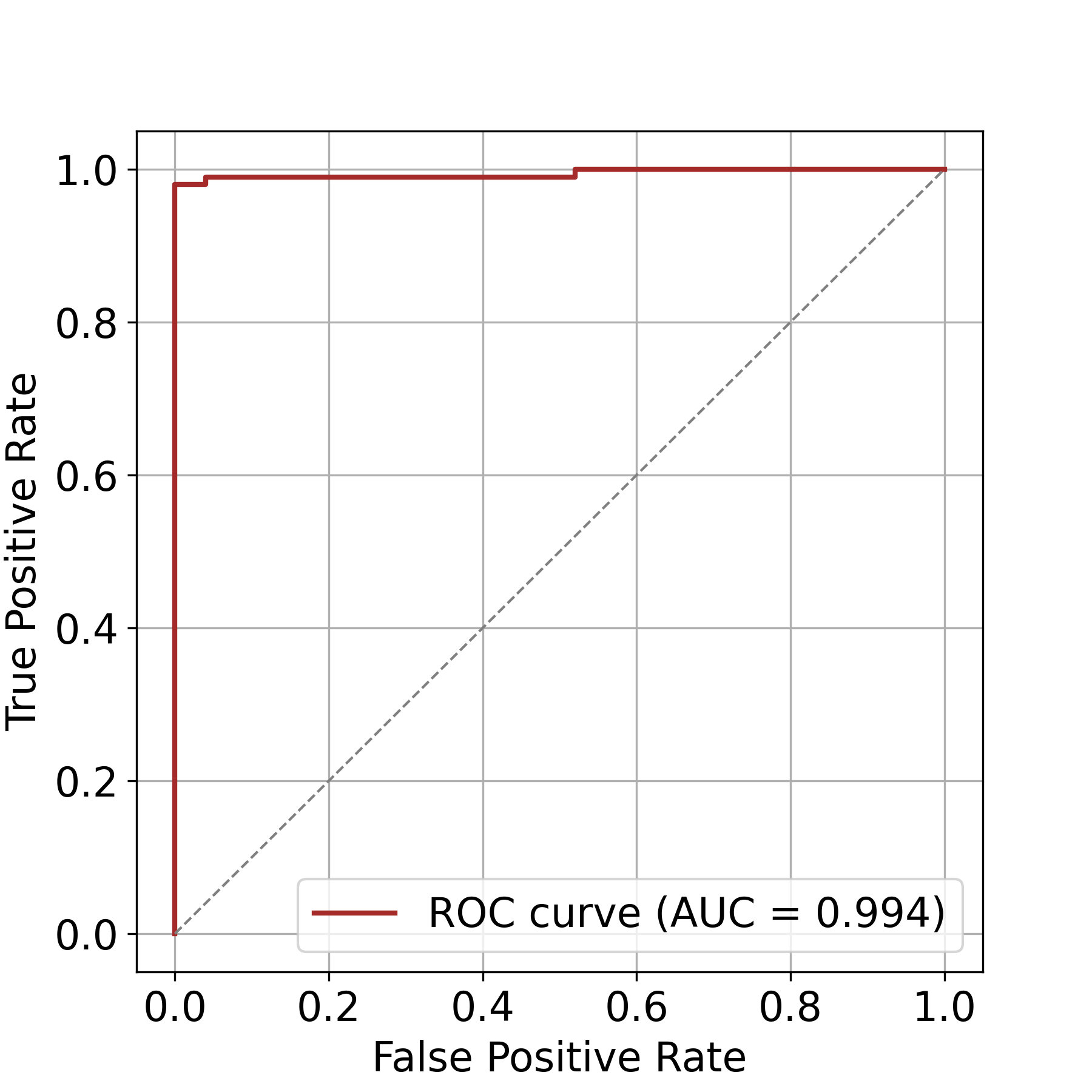}
        \vspace{0.1cm}
        (c)
    \end{minipage}
    
    \caption{Receiver Operating Characteristic (ROC) curves evaluating the classification performance of the architectures. Panels (a) corresponds to the result of training with Temperature maps, (b) U maps, and (c) Q maps.}
    \label{fig:ROCs}
\end{figure}

We provide additional validation of the architectures. \autoref{fig:ROCs} presents the Receiver Operating Characteristic (ROC) curves for the pipeline trained in a) the temperature maps, b) the U maps, and c) the Q maps. These networks were trained to differentiate between the targeted cosmological models directly using the CMB maps. As illustrated in the panels, all three architectures exhibit exceptional classification performance. The models achieve Area Under the Curve (AUC) scores of $0.996$, $0.983$, and $0.996$, respectively. These very high AUC values confirm that the networks possess a robust discriminative power and can successfully extract the necessary differentiating features directly from the map level.

\end{appendix} 

\end{document}